\documentclass[preprint,showpacs,preprintnumbers,amsmath,amssymb]{revtex4}

\usepackage{graphicx}
\usepackage{dcolumn}
\usepackage{bm}

\begin{document}

\title{\large Gating of Au-$\rm Bi_2Se_3$-NbN ramp-type junctions with superconducting NbN gate-electrode and $\rm SrTiO_3$ film as the gate-insulator }

\author{Gad Koren}
\email{gkoren@physics.technion.ac.il} \affiliation{Physics
Department, Technion - Israel Institute of Technology Haifa,
32000, ISRAEL} \homepage{http://physics.technion.ac.il/~gkoren}

\date{\today}
\def\bfig {\begin{figure}[tbhp] \centering}
\def\efig {\end{figure}}

\normalsize \baselineskip=8mm  \vspace{15mm}

\pacs{73.20.-r, 74.45.+c, 74.78.Fk, 74.90.+n }

\begin{abstract}

Ramp-type junctions of $\rm Au-Bi_2Se_3-NbN$ were prepared on top of a bottom gate comprised of a $\rm SrTiO_3$ gate-insulator film on $\rm NbN$ gate-electrode layer on (100) $\rm SrTiO_3$ wafer. Two wafers with gate-insulator thickness of 120 and 240 nm were studied, with the former showing higher gate leakage currents Ig at high gate voltages Vg, leading to heating effects and shifting of the junctions conductance spectra versus the voltage bias. At Vg=0 V, the conductance spectra of the low resistance junctions showed zero bias conductance peaks inside a tunneling gap with typical conductance drops when the critical current Ic was reached, while the high resistance ones exhibited tunneling conductance only. For Vg$>$-0.2 V ($\rm E\simeq$ -2 MV/cm) of the wafer with 120 nm thick gate-insulator linear Ig vs Vg was found, while for Vg$<$-0.2 V, Ig saturation was observed, leading to quadratic and linear heating effects at positive and negative high Vg values, respectively. This led to asymmetric conductance spectra shifts versus Vg which followed almost exactly the Ig vs Vg behavior. In the wafer with twice the gate-insulator thickness (240 nm), heating effects were strongly suppressed, and symmetric small peak shifts appeared only under the highest Vg values of Vg=$\pm$2 V ($\rm E\simeq \pm$ 10 MV/cm). Under Vg=2 V, a 5\% lower conductance was observed as compared to Vg=-2 V, indicating a small Fermi energy shift in our junctions under $\pm$10 MV/cm fields.

\end{abstract}

\maketitle

\section{Introduction }
\normalsize \baselineskip=6mm  \vspace{6mm}

Junctions of a normal metal (N) and a superconductor (S) with a topological insulator (TI) sandwiched in between may exhibit topological superconductivity (TOS). This can be a result of inducing superconductivity in the TI by means of the proximity effect with a conventional s-wave superconductor. TOS in the junctions in turn, can have zero energy modes of Majorana bound states (MBS), which exhibit Non-Abelian statistics and therefore might find use in future fault-tolerant quantum computers \cite{KaneRMP,Kitaev}. Gating of these junctions is essential in order to control their properties by tuning the Fermi energy of the TI to the Dirac point where the two surface bands cross inside the bulk energy gap \cite{KaneRMP,Steinberg,CMarcus}. Under these conditions, quantum operations by braiding of MBS are expected to be topologically protected, thus enabling basic quantum computations \cite{Kitaev,Nayak}. The present study is an extension of our previous reports on similar proximity systems comprised of $\rm Au-Bi_2Se_3-NbN$ junctions and $\rm Bi_2Se_3-NbN$ bilayers \cite{KorenPRB12,KorenEPL13,KorenSUST15,Koren17,KorenCondmat18}, but here we focus on gating effects of ramp-type junctions. To keep compatibility with standard fabrication processes, the junctions and gates are prepared from thin films only. By the use of NbN film as the gate-electrode and $\rm SrTiO_3$ (STO) as the gate-insulator, we find heating effects due to gate current leakage, conductance spectra shifts with gate voltage, and small Fermi energy tuning effect under a maximum field of about $\pm$10 MV/cm. \\

\section{Preparation and properties of the films and junctions }
\normalsize \baselineskip=6mm  \vspace{6mm}

The NbN, STO and $\rm Bi_2Se_3$ thin films were prepared by laser ablation deposition from a metallic Nb target or polycrystalline STO and $\rm Bi_2Se_3$ targets. While the NbN and STO films were deposited at 600 $^0$C heater block temperature under either 40 mTorr of $\rm N_2$ gas flow or vacuum, respectively, the $\rm Bi_2Se_3$ film were deposited at 300 $^0$C heater block temperature under vacuum. The corresponding wafer surface-temperatures during deposition were about 500 and 250 $^0$C, respectively. High laser fluence on the target was used for the deposition of NbN and STO ($\rm \sim 10\, J/cm^2$), while a much lower fluence was needed for the deposition of the $\rm Bi_2Se_3$ film ($\rm \sim 1\, J/cm^2$). Single layer films were deposited on (100) STO wafers of $\rm 10\times10\, mm^2$ area. X-ray diffraction measurements of these films showed that all grew with preferential crystallographic orientation. The NbN and STO layers, mostly in the cubic phase with a-axis orientation with a=0.433 nm for NbN and a=0.390 nm for STO, while the $\rm Bi_2Se_3$ film had the typical hexagonal structure with c-axis orientation normal to the wafer and c=2.88 nm. Metallic gold films were prepared also by laser ablation deposition under vacuum at $\rm \sim 10\, J/cm^2$ fluence on the target, but at 150 $^0$C heater block temperature.  \\

\begin{figure} \hspace{-20mm}
\includegraphics[height=8cm,width=11cm]{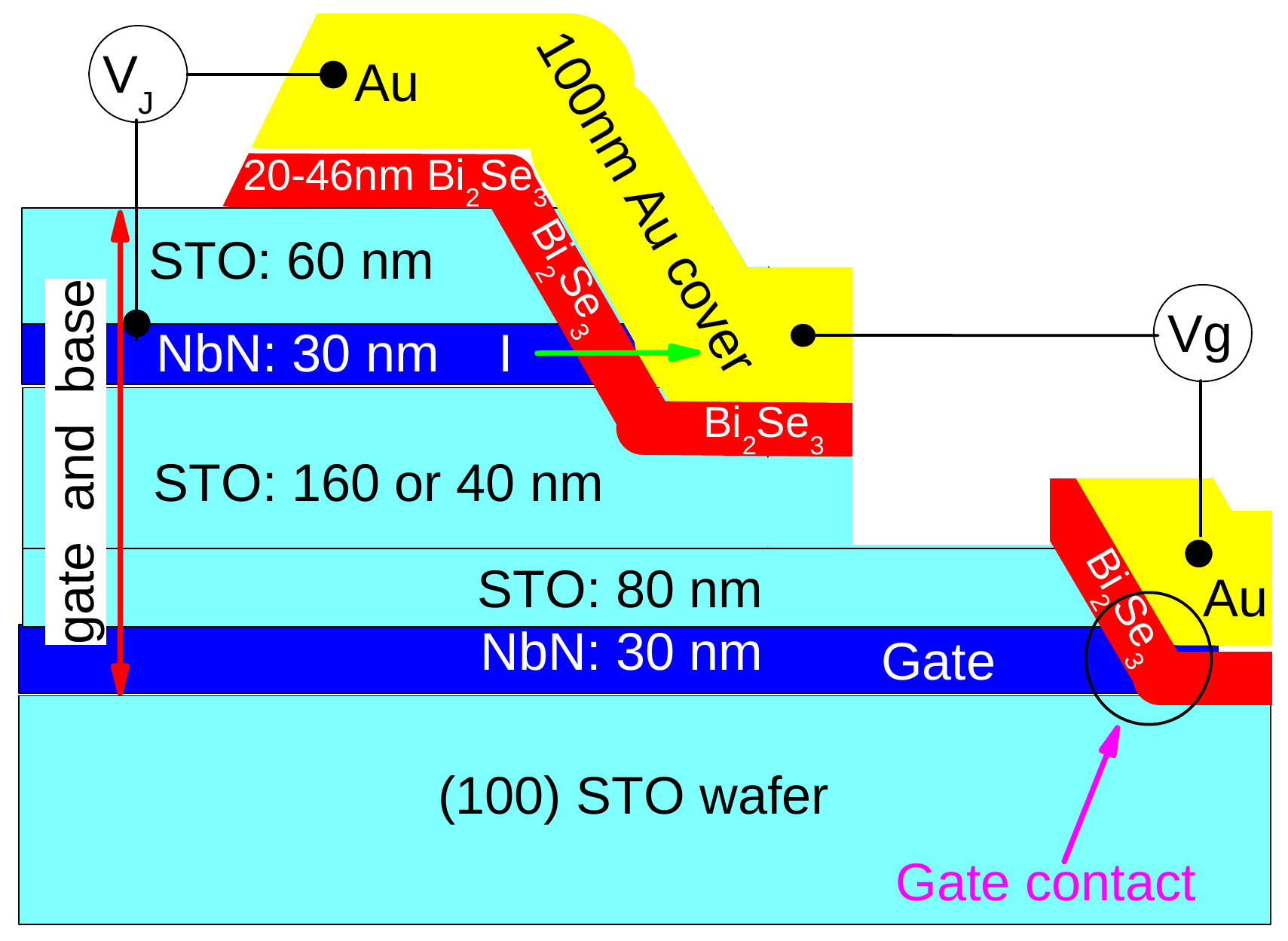}
\vspace{-0mm} \caption{\label{fig:epsart} (Color online) A schematic cross section of a ramp-type junction on top of a bottom gate comprising a 120 or 240 nm thick STO gate insulator film on 30 nm thick NbN gate-electrode. Top contact to the bottom gate is via a large area segment of the $\rm Au/Bi_2Se_3$ cover electrode separated from the rest of the junctions (right Vg contact), while the other contact is on top of the cover electrode. I is the bias current through the junction (green arrow) and $\rm V_J$ is the voltage measured on the junction.    }
\end{figure}

A schematic cross-section of the fully photo-lithographically patterned gate and ramp-type junctions is shown in Fig. 1. Similar edge (or ramp) type junctions are also described in detail in a previous study of our group \cite{Nesher}. First, two bilayers of the gate and base electrode are deposited in-situ in a multi step process without breaking the vacuum. The first bilayer comprises of the gate of 120 or 240 nm thick STO on 30 nm NbN, while the base bilayer is deposited on top of the gate and comprises of 60 nm STO on 30 nm NbN. During the deposition of the gate and after an 80 nm thick STO layer is already deposited, a metallic shadow mask is inserted on the edge of the wafer to allow for a top contact to the metallic gate-electrode (the bottom NbN layer, after milling of the ramp). Once the double bilayer is deposited, the wafer is taken out of the vacuum chamber and patterned by deep UV photolithography and Ar-ion milling to fabricate 10 base electrodes with their ramps and 20 contact pads for the junctions on half the wafer (see Fig. 2 (c) in Ref. \cite{KorenPRB12}). Then a cover electrode of 100 nm Au on 20-46 nm $\rm Bi_2Se_3$ is deposited on the patterned gate and base electrode, followed by a second patterning step to produce  10 separated junctions and 20 additional contact pads on the other half of the wafer. An Atomic force microscope image of a typical junction is shown in Fig. 5 (c). Conductance spectra were measured by the four-probe technique in the same cooling run on all the ten junctions, using a $4\times 10$ array of spring loaded gold coated tips pressed on the 40 gold contact pads on the wafer. \\

\section{Results and discussion}

\begin{figure} \hspace{-20mm}
\includegraphics[height=9cm,width=12cm]{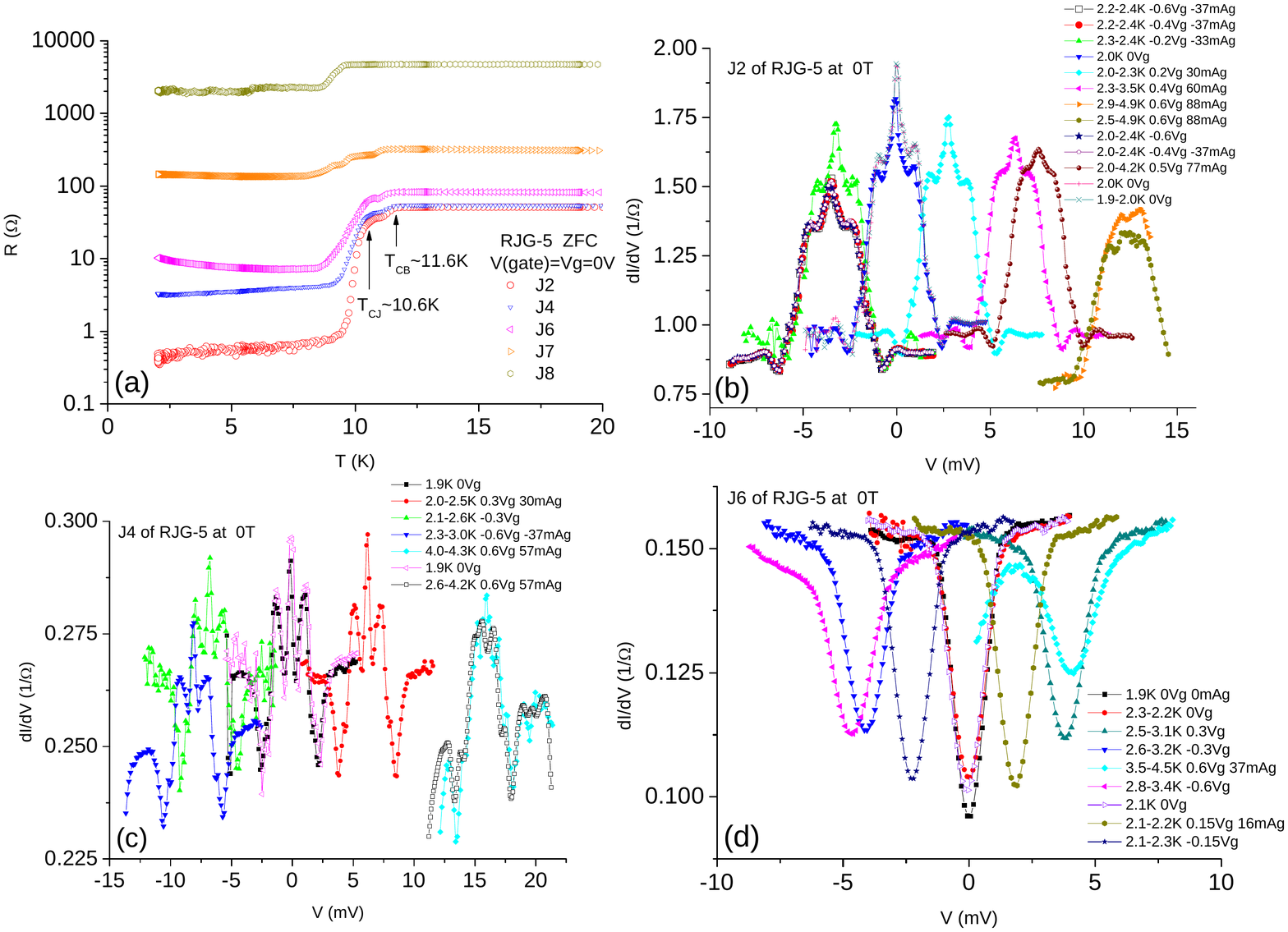}
\vspace{-0mm} \caption{\label{fig:epsart} (Color online) Transport results of junctions on the RJG-5 wafer with 120 nm thick STO gate-insulator and 20 nm thick $\rm Bi_2Se_3$ barrier. (a) Depicts the zero field cooled resistance vs temperature of five junctions on this wafer under Vg=0 V ("0Vg"). (b)-(d) Show conductance spectra of three junctions with the lowest resistances (J2, J4 and J6), under various gate voltages Vg (in V) and gate currents Ig (in mA, or "mAg" in short notation). All measurements in these panels started at a base temperature of about 2 K, but due to heating by the gate current, in particular at high positive Vg values, (see also Fig. 4), a range of temperatures is given for each spectrum.  }
\end{figure}

We start with the results obtained on the RJG-5 wafer which had an STO gate-insulator thickness of 120 nm and 20 nm thick $\rm Bi_2Se_3$ barrier. Fig. 2 (a) shows resistance vs temperature of five junctions on this wafer, obtained under zero field cooling (ZFC) and zero gate voltage (Vg=0 V or "0Vg" in short notation). One can see a large spread in the junctions resistance both in the normal state above the superconductive transition and below it. This is quite typical of ramp junction which are very sensitive to the cleanliness of the edge of the ramp regions on the wafer below the transition. The resistance spread above the transition can be due to thickness variations over the 10 mm width of the wafer. Similar but smaller spreads however, were also observed in our previous studies of all-epitaxial cuprate ramp junctions \cite{Aronov,KorenPRL11}. Thus the obviously non-epitaxial nature of the present junctions (hexagonal $\rm Bi_2Se_3$ on cubic NbN or STO can never be epitaxial) can also contribute to the large spread of the junctions properties. Nevertheless, the spread of resistances of the J2, J4 and J6 junctions is reasonable compared with previous studies, and we shall focus on the properties of these junctions. In Fig. 2 (a) one observes two superconducting transition temperatures for each of these junctions at about 11.6 and 10.6 K. We attribute the higher transition to the bulk NbN film of the leads to the junctions ($\rm T_{CB}$), and the lower one to the proximity effect (PE) transition of the junction itself ($\rm T_{CJ}$). Similar double resistance steps are typical of the bulk-leads and the junction-PE as observed previously in junctions and bilayers \cite{KorenSUST15,KorenLee16,KorenSUST17}. At low temperatures, J2 shows metallic behavior of the barrier, J6 insulating behavior, while J4 is somewhere in between with weak metallic decrease of resistance with decreasing temperature. These characteristics affect the conductance spectra of these junctions as seen in Fig. 2 (b-d) where we first focus on the 0Vg curves. J2 and J4 show clear zero bias conductance peaks (ZBCP) at low bias,  while the more resistive J6 exhibits tunneling-like behavior.\\

\begin{figure} \hspace{-20mm}
\includegraphics[height=9cm,width=13cm]{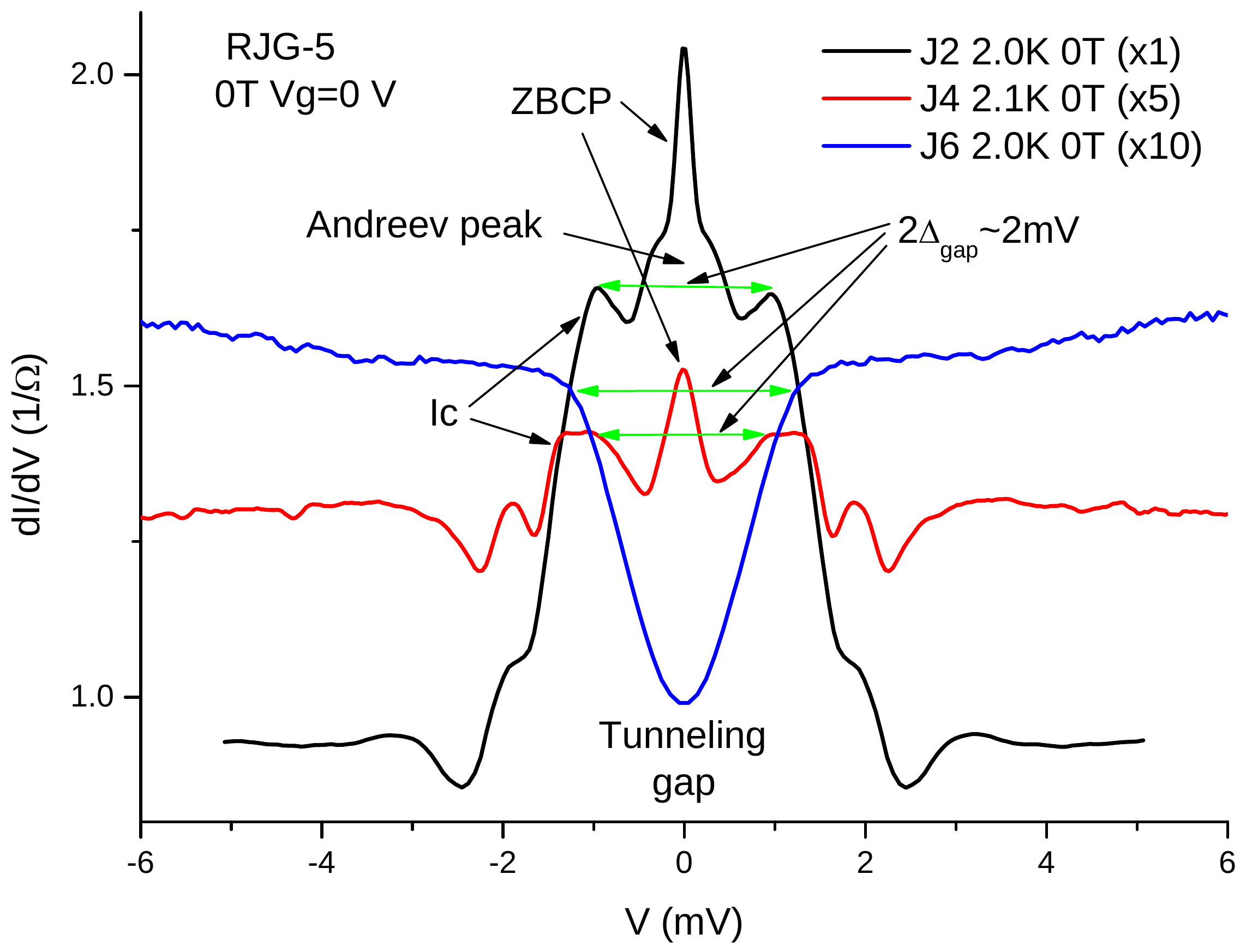}
\vspace{-0mm} \caption{\label{fig:epsart} (Color online) Conductance spectra of three junctions of Fig. 2 under 0T and 0Vg. Note the different multiplication factors used to present clearly all spectra on the same panel.}
\end{figure}

Before discussing the effects induced by applying different gate voltages as in Fig. 2 (b-d), we replot in Fig. 3 three conductance spectra of these J2, J4 and J6 junctions under 0Vg, but with different multiplication factors to enable clear presentation and comparison between them on the same panel. We already mentioned the ZBCPs seen in the spectra of J2 and J4, and here we see that they are embedded in a gap-like structure where J2 also exhibits two clear coherence peaks at about $\pm$1 mV. J6 on the other hand shows tunneling behavior only, with energy gap $\Delta$ also of about 1 meV but with no coherence peaks. Both, the ZBCPs and tunneling phenomena are due to Andreev reflections and Andreev bound states in junctions with different transparencies as discussed in detail in Refs. \cite{BTK,MilloKoren17}. At higher voltage bias in the low resistance (high transparency) junctions J2 and J4 sharp conductance drops are observed at 1.1 and 1.4 mV. These are attributed to reaching the critical current $\rm I_c$ in the narrowest lead to the junctions, they are followed by dips (clearly seen for J4) and are due to well known heating effects \cite{Sheet}. Also seen in Fig. 3 are a broad Andreev peak inside the gap of J2, and some above-gap peaks in J2 and J4 which could originate in other types of bound states.\\

\begin{figure} \hspace{-20mm}
\includegraphics[height=9cm,width=13cm]{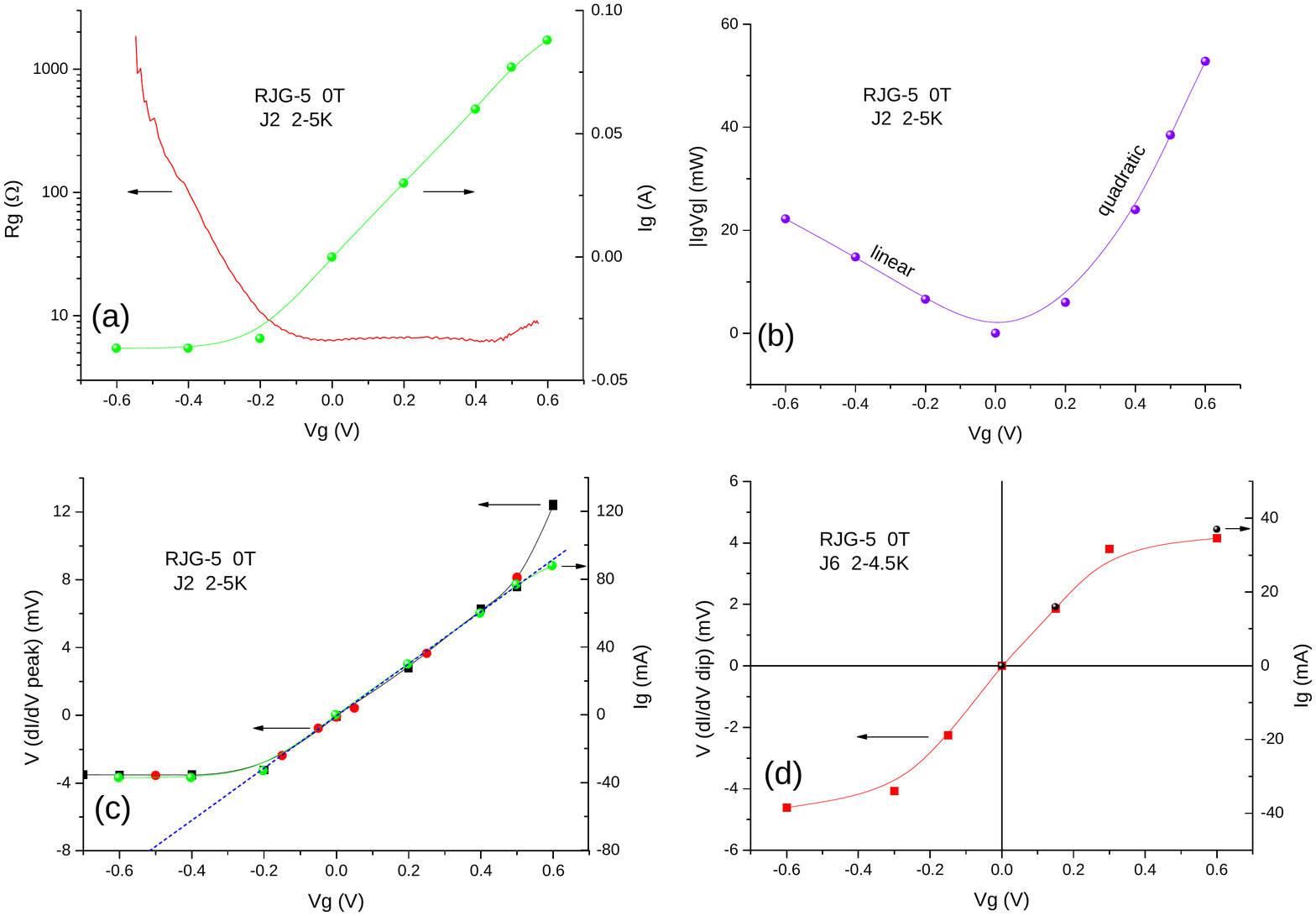}
\vspace{-0mm} \caption{\label{fig:epsart} (Color online) (a) Depicts gate resistance Rg and gate current Ig vs Vg and (b) shows the heating power $\rm |IgVg|$ vs Vg for the wafer of Fig. 2. In (a), the rectification-like saturation of Ig vs Vg below Vg=-0.2 V is a result of the sharply increasing Rg at these voltages. (c) and (d) show the voltages of the peaks and dips of the conductance spectra of the J2 and J6 junctions, respectively, vs Vg. These follow Ig almost exactly except for the data point at Vg=0.6 V in (c) which is a result of extra heating (see also Fig. 2 (b)).    }
\end{figure}

\begin{figure} \hspace{-20mm}
\includegraphics[height=9cm,width=11cm]{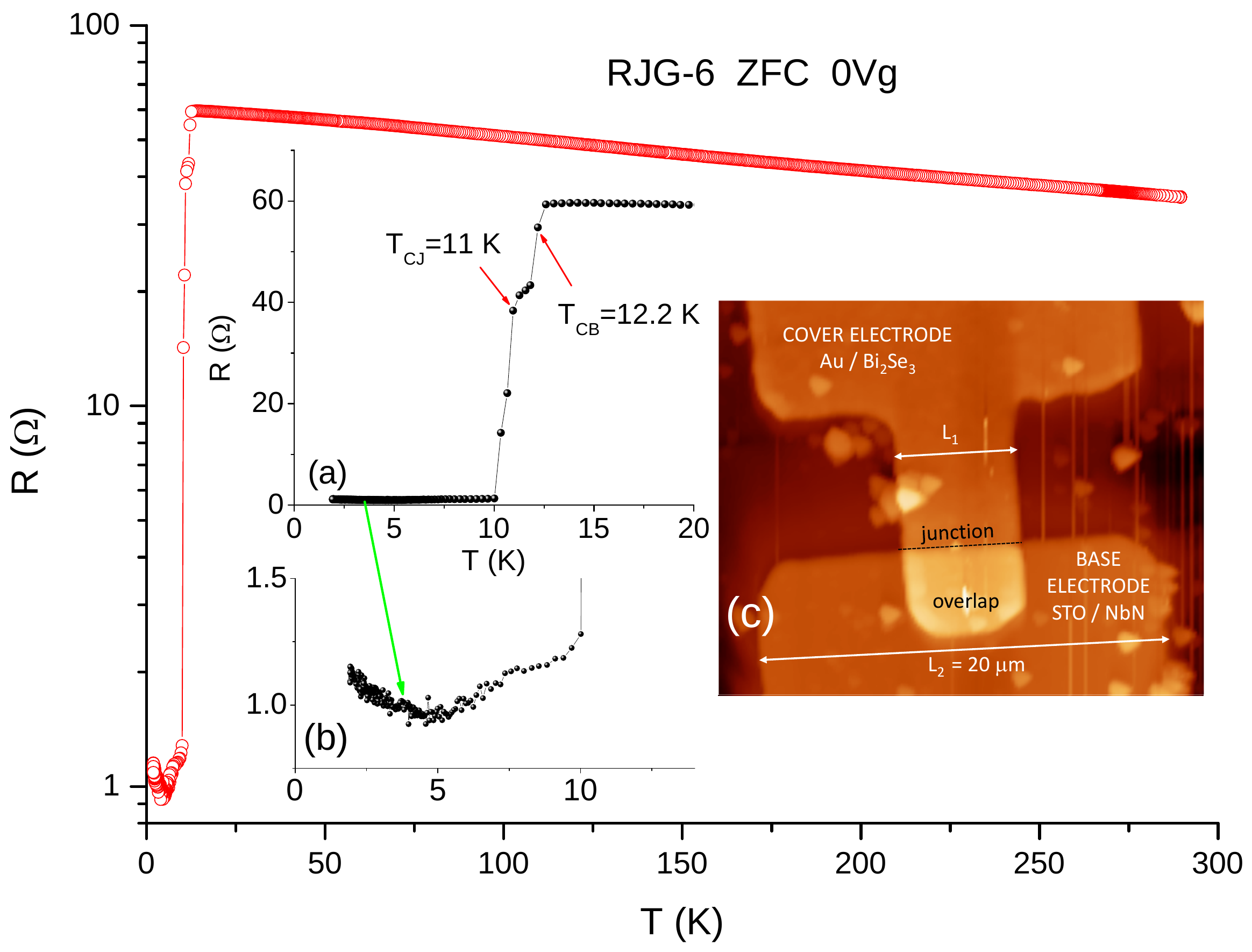}
\vspace{-0mm} \caption{\label{fig:epsart} (Color online) Measured results on the RJG-6 wafer with 240 nm thick gate-insulator and 46 nm thick $\rm Bi_2Se_3$ barrier. ZFC resistance versus temperature under 0Vg. Inset (a) depicts the transition regime showing the bulk NbN base electrode transition at $\rm T_{CB}$=12.2 K and the junction proximity effect transition at $\rm T_{CJ}$=11 K. Inset (b) is a zoom-in on the low temperature regime, and inset (c) shows an atomic force microscope image of the ramp-type junction. The junction is located under the dotted line and its inclined cross-sectional area is of about $\rm 6\times 0.5\,\mu m^2$. Note that no current can flow normal to the wafer from the base to cover electrode in the overlap area due to the insulating STO layer in between the two.
 }
\end{figure}

We now go back to discuss the gate voltage effects observed in Fig. 2 (b-d). The immediate observation is that all conductance spectra are shifted with the applied voltage on the gate Vg while basically keeping their symmetric shape. It is also worth noting that these spectra shifts are in the few mV regime while the applied gate voltages are of about two orders of magnitude higher. To understand this behavior, we plot in Fig. 4 various parameters of the gate and the shifts of the the conductance spectra peaks or dips vs Vg. Fig. 4 (a) shows the dynamic gate resistance  Rg and gate current Ig  of J2 vs Vg. A clear asymmetry of the response is observed between applying positive or negative gate voltages Vg. While an almost linear Ig vs Vg is found for Vg$>$-0.2 V, below this gate voltage there is a tendency to saturation. As a result, the corresponding gate resistances are almost constant above this Vg, and increasing sharply below it. This leads to a quadratic and linear power dissipation $\rm |IgVg|$ in the junction for positive and negative Vg values, respectively, as seen in Fig. 4 (b). Thus gate heating effects are more pronounced under positive Vg, in particular at high Vg values. In Fig. 4 (c), we plot the voltages of the peaks of the spectra of Fig. 2 (b) on the same graph of Ig vs Vg of Fig. 4 (a), but with a scaling factor to make them coincide at low Vg. We find that both set of data have the same behavior for all Vg values, except for the peak data point at Vg=0.6 V which deviates sharply from the Ig graph, due to extra heating under this Vg value (see the higher temperature in Fig. 2 (b) under Vg=0.6 V). Fig. 4 (d) basically shows similar results at least up to Vg=0.3 V, but we have no sufficient data for this J6 junction to draw further conclusions. We conclude from the results of Fig. 4 that our gate insulation is quite poor and leads to high gate leakage current Ig at high Vg. The fact that the spectra shifts in Fig. 2 (b-d) vs Vg follow almost exactly those of Ig vs Vg, indicates that they are mainly the result of the gate leakage current Ig.\\

\begin{figure} \hspace{-20mm}
\includegraphics[height=8cm,width=11cm]{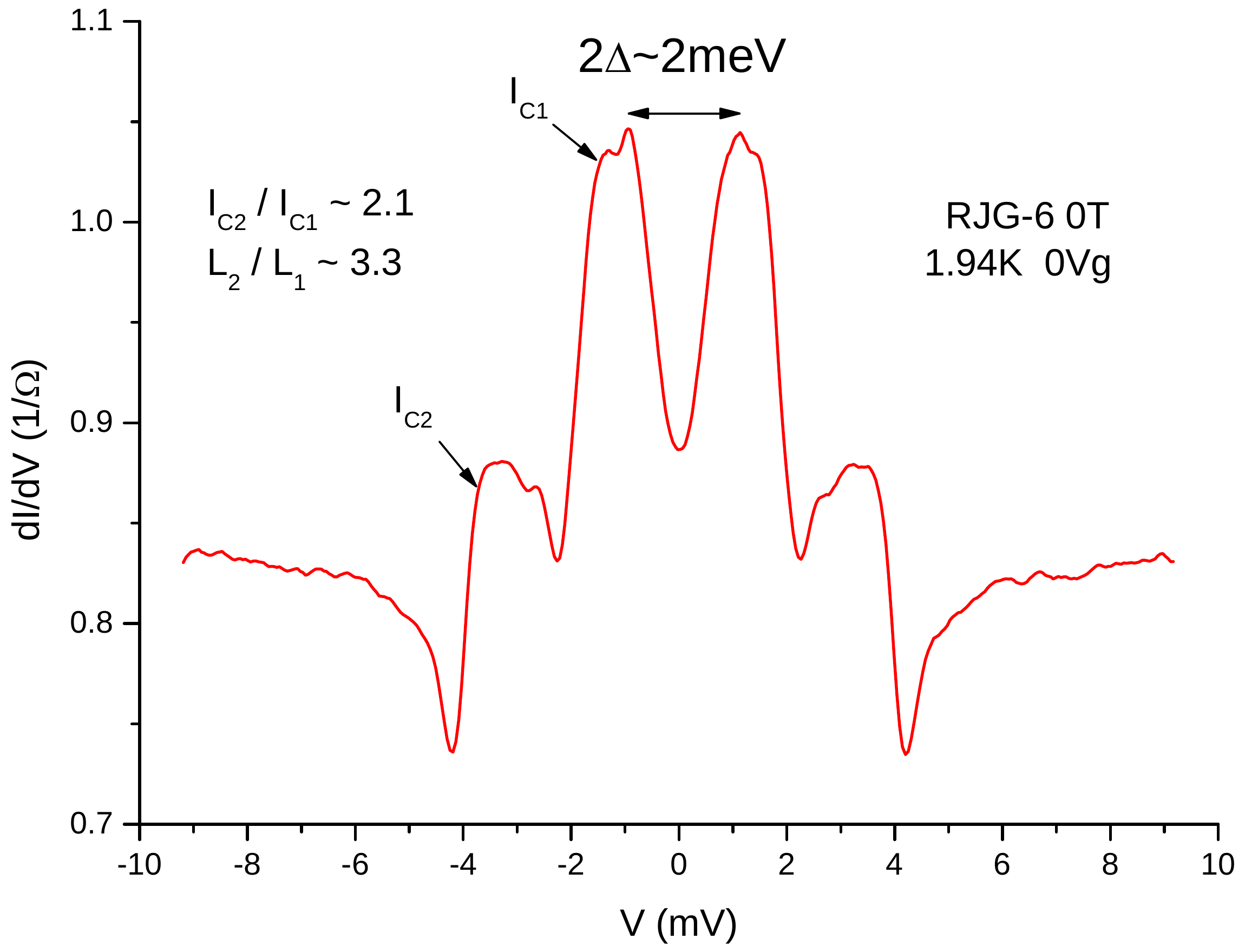}
\vspace{-0mm} \caption{\label{fig:epsart} (Color online) Typical conductance spectrum at 1.94 K of  a junction on the wafer of Fig. 5 under 0T and Vg=0 V. Note that the junction resistance here is of about 1 $\Omega$, similar to that of J6 in Fig. 3, thus the main feature at low bias is tunneling with critical currents $\rm I_c$-drops and dips at higher biases. $\rm I_{ci}$  with i=1 and 2 are defined as $\rm \rm V_i\times (dI/dV)_{V_i}$ while the ratio $\rm L_2/L_1$ is taken from the image of Fig. 5.  }
\end{figure}

To avoid this unwanted spurious behavior of the junctions due to the leaky gate, we prepared another wafer RJG-6 with a thicker gate insulating STO layer of 240 nm thickness and 46 nm thick $\rm Bi_2Se_3$ barrier. Resistance vs temperature results under ZFC and 0Vg are depicted in the main panel of Fig. 5. Inset (a) shows the transition regime with two transitions at $\rm T_{CB}$=12.2 K and $\rm T_{CJ}$=11 K, of the bulk NbN lead to the junction and the junction itself due to the proximity effect, respectively, similar to the results in Fig. 2 (a). Inset (b) is a zoom-in on the low temperature R vs T which shows that R below the PE transition is not constant. Inset (c) shows an atomic force microscope image of this junction, with the base and cover electrodes as marked in this image, and the junction location along the dotted line where the base and cover electrodes are connected at the ramp as seen in Fig. 1. The conductance spectrum of this junction at low temperature and 0Vg is shown in Fig. 6. It shows a clear tunneling behavior at low bias, with two coherence peaks and energy gap $\Delta$ of about 1 meV,  similar to that of J6 of Fig. 3 which has a similar junction resistance of about 1 $\Omega$, but no coherence peaks. In addition and unlike J6 of Fig. 3, the present junction exhibits also two sharp conductance drops when two critical currents ($\rm I_{c1}$ and $\rm I_{c2}$) are reached, followed by the typical $\rm I_{ci}$ dips \cite{Sheet}. Using the critical current definition: $\rm I_{ci} \equiv \rm V_i\times (dI/dV)_{V_i}$, we find that $\rm I_{c2}$/$\rm I_{c1} \sim 2.1$ (where $\rm I_{c1}$ corresponds to the base electrode and $\rm I_{c2}$ to the cover electrode) while $\rm L_2/L_1\sim 3.3$ as obtained from the image of Fig. 5 (c). For uniform current distribution in the current leads to the junction, these two ratios should be equal. The fact that they are quite different, is due to a non-uniform current distribution where the current tends to flow along the edges of the superconducting leads.\\

\begin{figure} \hspace{-20mm}
\includegraphics[height=8cm,width=11cm]{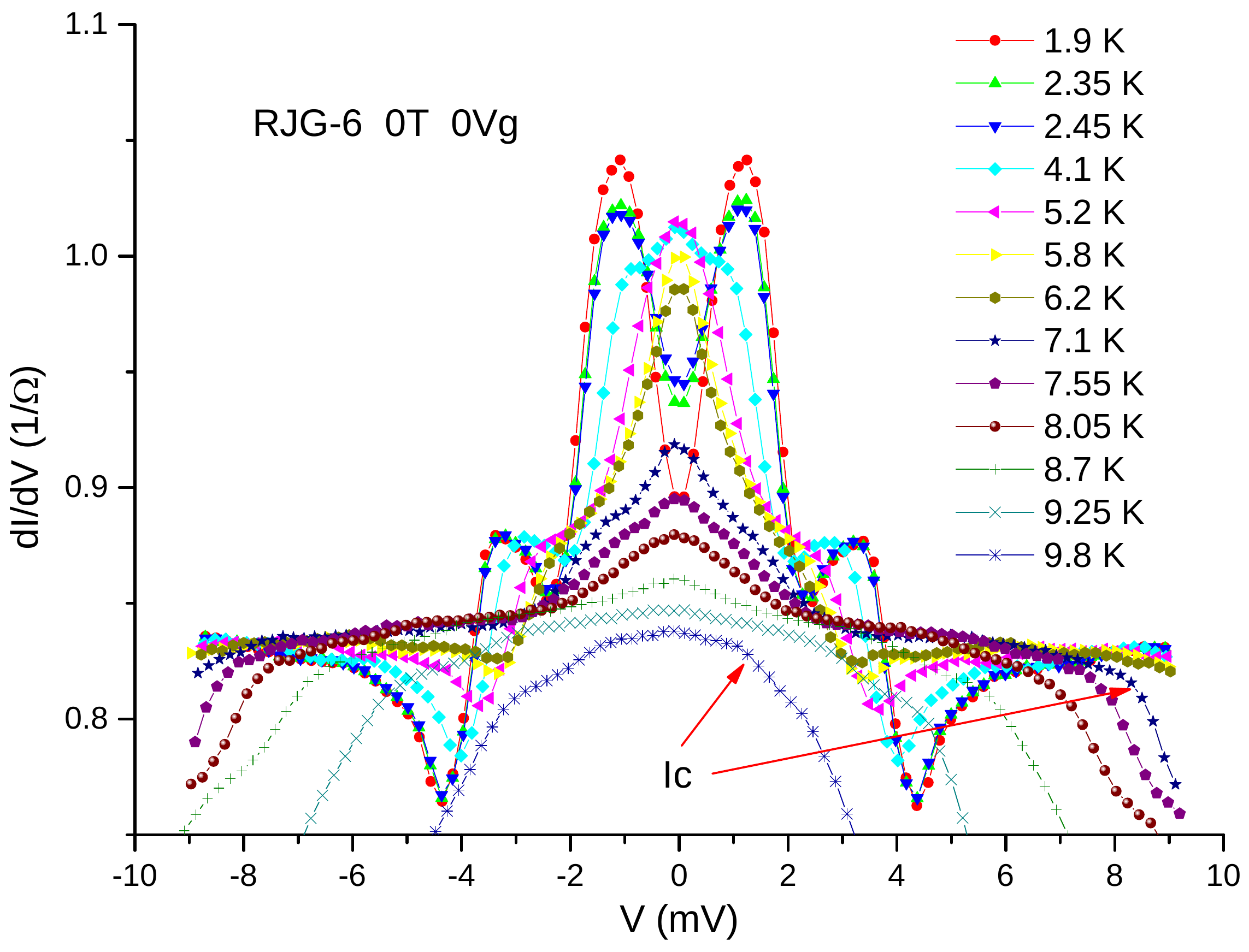}
\vspace{-0mm} \caption{\label{fig:epsart} (Color online) Conductance spectra at various temperatures of the junction of Fig. 5 under 0T and 0Vg.
 }
\end{figure}

Since heating effects by the gate-leakage currents Ig were found for junctions on the RJG-5 wafer as seen in Figs. 2-4, it is important to know how the spectra look like at different temperatures. We therefore measured the conductance spectra at various temperatures of the junction of Figs. 5 and 6 on the RJG-6 wafer under 0Vg and 0T as shown in Fig. 7. Here the temperatures were stable with no drifts during each measurements. One can see that by raising the temperature, the tunneling gap fills up and a single ZBCP is formed which is decreasing in magnitude with further increase of temperature. Similarly, the sharp $\rm I_{ci}$ drops shift to low bias with increasing temperature, and new ones formed at high bias due to wider NbN lead-segments (not shown in the image of Fig. 5) that reach $\rm I_{c}$ at even higher temperatures, until they also vanish at $\rm T_{C}$ of about 10 K.\\

\begin{figure} \hspace{-20mm}
\includegraphics[height=8cm,width=11cm]{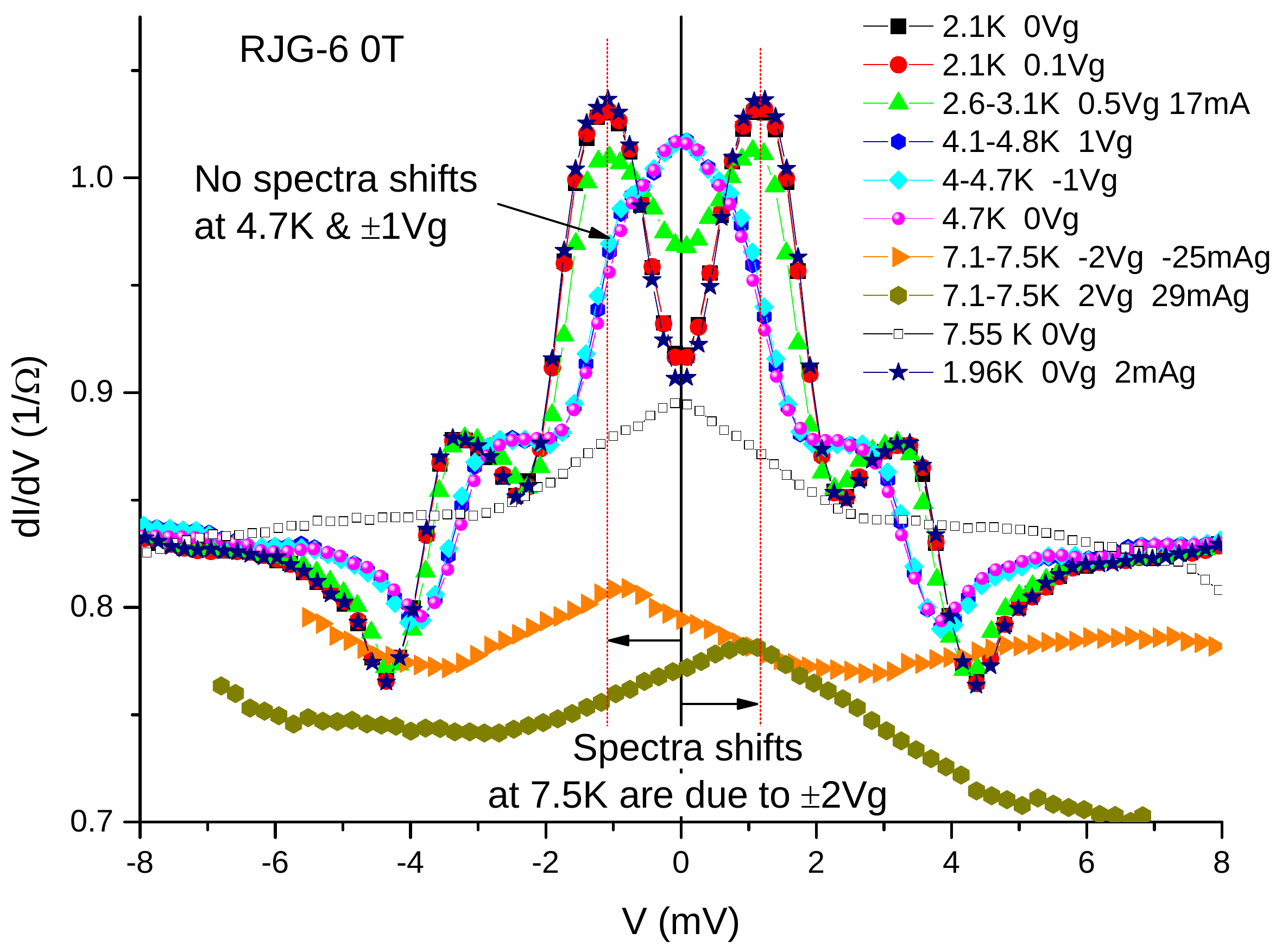}
\vspace{-0mm} \caption{\label{fig:epsart} (Color online) Conductance spectra of the junction of Fig. 5 under various gate voltages. All measurements started at low temperature of about 2 K but due to heating effects by the gate at high Vg values, the temperature increased during the measurements, and a temperature range is therefore given for these spectra. Two spectra at stabilized 4.7 and 7.55 K both under 0Vg are shown for comparison with the spectrum under $\pm$1Vg at 4-4.7 K and the spectrum under $\pm$2Vg at 7.1-7.5 K.
 }
\end{figure}

Next the conductance spectra of the junction on the RJG-6 wafer under different gate voltages Vg are presented and discussed. The resulting spectra are depicted in Fig. 8, and in all of them the measurement started at low temperature of about 2 K (the base temperature). If heating occurred during the recording process (typically 30-60 s), then a temperature range is given in this figure as indicated. One can see that no heating occurs up to 0.1Vg as T remains 2.1 K, but a small heating occurs 0.5Vg  to 2.6-3.1 K, which increases at $\pm$1Vg to 4.1-4.8 K. Nevertheless, unlike in Fig. 2 (b-d) the observed spectra do not shift here as compared to the 0Vg spectrum at the same temperature (4.7 K and 0Vg). Under $\pm$2Vg however, the highest gate voltages applied here which correspond to about $\pm$10 MV/cm gate field for $\epsilon \sim 120$ \cite{Mannhart} (or $\pm$25 MV/cm if $\epsilon \sim 300$ is use \cite{Venki}), significant heating to 7.1-7.5 K occurred. This resulted in spectra  with slightly asymmetric gate currents Ig of -25 mA at -2Vg and 29 mA at 2Vg. In this case, the spectra did shift by a small voltage to about $\pm$1 mV, and also had lower maximum conductance as compared to the spectrum taken at 7.55 K and 0Vg (no gate heating). This comparison also indicates that the shape of the spectra under $\pm$2Vg remained almost the same as under 0Vg at the same temperature. We conclude that the gate heating effects are much smaller in the junction on the RJG-6 wafer as compared to those on the RJG-5 wafer, but still exist. Although bulk single crystals of STO are excellent insulators with very high dielectric constant at low temperatures ($\epsilon \sim 10^4$) \cite{Muller}, in the form of sputtered or laser ablated thin films, STO is a much poorer insulator due to many structural defects and trapped charges  ($\epsilon \sim 120-300$) \cite{Mannhart,Venki}. The fact that the peak conductance at 2Vg is lower by about 5\% than that at -2Vg can be due to real Fermi energy tuning as one expects from effective gating, but this effect is quite small. Part of it can still be due to the slightly higher Ig under 2Vg as compared to -2Vg, which would lead to higher resistance (and lower conductance) near 7.5 K due to increasing temperature as seen in Fig. 5 (b). Possibly, the applied gate-field was not sufficiently high to induce a significant Fermi level shifting or tuning effect in the $\rm Bi_2Se_3$ barrier layer of the ramp-junctions. \\

Finally, we note that in Ref. \cite{Venki} the use of an Au gate-electrode yielded a higher breakdown field than with a Pt gate-electrode. This is apparently due to the higher conductivity of Au which is about 4-5 times higher than that of Pt. Therefore, the use of a superconducting gate-electrode with zero resistance in the present study should have yielded an even higher breakdown voltage. However, the present gate area (including the contacts) is about 5 mm$^2$ per junction as compared to 0.3 mm$^2$ in Ref. \cite{Venki}. Thus chances of having a strongly leaky defect in our gate-insulator are about 16 times higher. Moreover, the second gate electrode in the present study is the gold layer of the cover electrode. Hence it is uncertain by how much our breakdown voltage was higher than in \cite{Venki}. \\

\section{Conclusions}

On-wafer gating of ramp-type junctions of $\rm Au-Bi_2Se_3-NbN$ with superconducting NbN gate-electrode and $\rm SrTiO_3$ gate-insulator is demonstrated. Heating effects and conductance spectra shifts due to gate-leakage currents are stronger with  a thinner gate-insulator layer than with a thicker one. A small Fermi level tuning of 5\% is observed in the $\rm Bi_2Se_3$ barrier layer of the ramp-junctions under $\pm$10-25 MV/cm gate field for $\epsilon \sim 100-300$.


\bibliography{AndDepBib.bib}

\bibliography{apssamp}

\end{document}